\documentclass[preprint,prc,aps,showpacs,showkeys,groupedaddress,floatfix]{revtex4-2}
\usepackage{epsfig}
\usepackage{dcolumn}
\usepackage{bm}
\usepackage{graphics}
\usepackage{graphicx}
\usepackage{subcaption} 
\begin{document}

\title{Exploring Anomalous Flavor-Changing Neutral tqh Transitions at Future Muon Colliders: Insights from $\mu^{+}\mu^{-}\rightarrow t\bar{q}h$
Interactions. }

\author{E. Alici}
\email[]{edaalici@beun.edu.tr} \affiliation{Department of Physics, Zonguldak Bulent Ecevit University, Turkey}

\begin{abstract}
Flavour-changing neutral current (FCNC) interactions between the Higgs boson and the top quark are subject to significant suppression within the Standard Model (SM). Consequently, these interactions can only attain an observable level through the influence of Beyond the Standard Model (BSM) physics effects. It is therefore of great importance to investigate these rare interactions, both to test the limits of the Standard Model and to reveal new physical signatures. At this point, the high sensitivity, clean experimental environment and low systematic uncertainties provided by muon colliders allow for a detailed study of such rare processes.

In this context, the present study analyses the signal of the process $\mu^+\mu^- \to t\bar{q}h$ and the background processes $t\bar{t}$, $tW\bar{b}$, $WWh$ and $WW$ at both in $\sqrt{s} = 1$ TeV and $\sqrt{s} = 3$ TeV energy levels. Analyses at the $1 TeV$ energy level reveal the observability of FCNC interactions even at low energy conditions and achieve the limit $\text{Br}(t \to qH) = 1.96 \times 10^{-4}$. This result corresponds to an improvement of approximately 1.5 times compared to the current limits of the ATLAS experiment, which are $3.0 \times 10^{-4}$. 
On the other hand, at the 3 TeV energy level, the most stringent limit was calculated under the luminosity $\mathcal{L} = 20 ab^{-1}$ ,  yielding a value of  $\text{Br}(t \to qH) = 1.1 \times 10^{-4}$. This result coincides an improvement of approximately 3 times with respect to the current ATLAS experimental limits. 
The findings of the study underscore the potential of muon colliders in the search for new physics and the pivotal role they can play in the study of rare processes such as Higgs-top quark interactions. Consequently, the results are expected to provide a crucial foundation for both experimental and theoretical studies and to make a significant contribution to the search for BSM physics.


\end{abstract}


\maketitle

\section{Introduction}

The Standard Model (SM) is one of the most successful theoretical frameworks in elementary particle physics and provides an explanation of the fundamental forces of nature and their interactions. The SM unifies the electromagnetic, weak and strong nuclear forces on the basis of quantum field theory. It has also demonstrated considerable success in explaining processes such as the Higgs mechanism\cite{hh,hh1}. Furthermore, the SM has been repeatedly confirmed by experimental observations, thereby becoming one of the milestones of our understanding of fundamental physics at today.
However, the scope of the SM is constrained, and it is inadequate to account for all phenomena observed in the universe. For instance, while gravity cannot be described within the scope of the SM, it also cannot provide any explanation for cosmological phenomena such as dark matter and dark energy. Similarly, theoretical problems such as the masses of neutrinos and the hierarchy problem regarding the mass of the Higgs boson indicate that the SM is an incomplete theory. In this regard, the search for new theories of physics beyond the SM (BSM) has become a significant priority in the study of contemporary physics.
At this juncture, an effective approach to understanding physics beyond the SM can be provided by the Standard Model Effective Field Theory (SMEFT) framework. SMEFT is a powerful tool that extends the SM at low energy limits and allows to analyse new physics effects in an energy scale independent manner. In this framework, the effects on the SM of new particles or interactions thought to exist at high energy scales can be effectively expressed. In summary, SMEFT provides a flexible framework for both theoretical and experimental research and serves as a bridge to new physics beyond the SM.
On the other hand, one of the most crucial processes investigated to ascertain the boundaries of the SM is that of Flavour-Changing Neutral Current (FCNC) interactions. At the SM, these interaction processes exclusively occur at the loop level and are significantly suppressed \cite{1}. These processes provide an ideal means of tracing interactions beyond the SM through precise experimental measurements. In particular, in the case of heavy fermions such as top quarks ($t$), FCNC processes are extremely sensitive to new physics.
In this context, the process focussed in the present work is the $t \to qh$ ($q = u, c$) transition, which is one of the FCNC interactions with the Higgs boson ($h$). In the SM framework, the probability of this process is extremely low ($\text{Br}(t \to qh) \sim 10^{-15}$)\cite{2,3}. However, extended theoretical frameworks such as SMEFT provide a window to observe the beyond-SM physics signatures associated with such processes as mentioned above. In particular, new FCNC interactions between the Higgs boson and the top quark can be parametrically analysed via SMEFT operators, and the experimental limits of these operators are also being tested at existing colliders such as the LHC \cite {17,18,19,20,21,22,23}. The most recent experimental work in this area was conducted by the ATLAS collaboration using $140 fb^{-1}$ data collected at the 13 TeV energy level. In this experiment, the $t \to qh$ processes through the two-photon decay of the Higgs boson ($H \to \gamma\gamma$) channel were investigated and upper limits for branching ratios were found to be $\text{Br} < 3.0 \times 10^{-4}$ for $t \to cH$ and $\text{Br} < 3.8 \times 10^{-4}$ (95\% confidence level) for $t \to uH$ \cite {23}. These results once again demonstrate the potential of FCNC processes to probe new physics beyond the SM.
It is at this juncture that the significance of phenomenological studies becomes apparent.\cite{4,5,6,7,8,9,10,11,12,13,14,15,16} Phenomenology serves to establish a connection between theoretical predictions and experimental data, thereby facilitating both the experimental assessment of theoretical models and guiding future experimental designs. Phenomenological analyses, particularly within the context of model-independent approaches such as SMEFT, contribute to the determination of parameter ranges that can be measured in experiments. This allows the experimental observation framework in which to search for signs of physics beyond the SM to be narrowed and more focused analyses to be performed. 
In this context, next-generation experimental platforms play a vital role for the verification of phenomenological predictions and the discovery of signals beyond the SM. Phenomenological studies on various types of colliders have clearly demonstrated the potential of these platforms in the study of BSM processes. For example, proton-proton colliders provide high event counts in rare processes related to the Higgs sector and heavy quarks, enabling the search for new physics. However, complex hadronic backgrounds can make precise measurements at such colliders difficult. Electron-positron colliders, with their low background levels, offer distinctive advantages in precision measurements and have been used to test anomalies related to the Higgs boson. However, due to energy density limitations, these colliders have difficulties in the production of heavy particles. At this point, muon colliders stand out as a unique platform for the study of rare processes such as FCNC, thanks to their high energy density and their structure free from hadronic backgrounds.
Muons belong to the lepton class, which also includes electrons. Although they exhibit similar properties in terms of charge and spin, muons are approximately 200 times heavier than electrons. The higher mass enables the release of greater energy density during the collision, thus increasing the probability of observing heavy particles and rare processes.  In the muon colliders compared to electron colliders, most of the energy also convert to mass and the synchrotron radiative loss is much lower. These characteristics enable them for more efficient collisions at high energy levels. Furthermore, the occurring collisions in muon colliders unlike proton ones are ‘clean’, in other words, the collisions are occurring among non-compound elementary particles, which makes events easier to analyse. Hence, this situation allows for a more precise study of processes beyond the SM. In particular, the study of rare processes such as $\mu^{+}\mu^{-}\rightarrow t\bar{q}h$ requires high energy and precision, and at this juncture, muon colliders provide an ideal platform for such experiments as previously stated reasons. In a summary, it can be said that the muon colliders become a significant tool to extend the observability limits of rare BSM processes, such as FCNC transitions.
In the light of this important literature given above, $\mu^{+}\mu^{-}$ The process $t \to t\bar{q}h$ is analysed in the SMEFT framework by considering the transitions $t \to qh$ ($q = u, c$) and the new physics potential of this process in a muon collider is evaluated. In this context, the calculations performed to evaluate the potential of muon colliders were conducted at $\sqrt{s} = 1$ TeV and $\sqrt{s} = 3$ TeV centre-of-mass energy.  These energy values correspond to the optimum operating range provided by muon colliders, both to go beyond current theoretical limits and to improve the observability of rare processes. Moreover, the effective luminosity range of the muon beam in our analyses were chosen as $1-20ab^{-1}$, in line with the annual luminosity targets proposed by the Muon Collider Collaboration, and thus we aimed to guarantee the experimental feasibility of our analyses\cite{24,25}. As a result, the ultimate goal of the work is to go beyond the limits of the SM, to shed light on the nature of FCNC processes and to reveal more precisely the signs of new physics.

\section{Theoritical Framework}\label{sec3}
In this paper, we adopt a phenomenological approach to examine the process $t \to qh$ ($q = u, c$) in search of physics beyond the SM. As previously outlined, the SMEFT framework both preserves the assumptions of the SM and offers the possibility of modelling new physics effects through higher order operators. Consequently, this framework allows to explore the traces of new physics that may arise at energy scales that are not experimentally accessible, by expressing them in terms of parameters effective at low energy limits.
SMEFT represents a particularly powerful tool for the study of rare processes. It is well established that the rarely observed phenomena, such as FCNC interactions between the Higgs boson and the top quark, are strongly suppressed within the SM. However, these processes offer a distinctive opportunity to explore new physics due to their sensitivity to high energy physics in the SMEFT framework. Therefore, rare decays like $t \to qh$ are one of critical processes that need to be studied both theoretically and experimentally.
In this context, the theoretical foundation of the study is based on an extended Lagrangian that describes Higgs-top quark interactions and FCNCs. This Lagrangian includes higher-order operators representing new physics effects while it preserves the basic assumptions of the SM.  The anomalous $(t \to Hq),(q = u, c)$ transitions between the Higgs boson and top quark and Higgs-fermion interactions are particularly modelled by the following effective interaction term\cite{29}:

\begin{eqnarray}
{\mathcal{L}}={\kappa}_{tuH}\bar{t}Hu+{\kappa}_{tcH}\bar{t}Hc+h.c.
\end{eqnarray}

Here, $\kappa_{tuH}$ and $\kappa_{tcH}$ represent the parameters of the FCNC interactions between the Higgs boson and the top quark. In the study, it was assumed that $\kappa_{tqH} = \kappa_{tqH}^\dagger = \kappa_{qtH} = \kappa_{qtH}^\dagger$ ($q = u, c$) and, these parameters are symmetric and real.
This form of the Lagrangian enables modelling of direct interactions between the Higgs boson and the top quark, thereby overcoming the limitations of the SM. Hence, this extended Lagrangian provides a fundamental theoretical framework for the investigation of the $t \to qh$ process.
Also, the branching ratio of the $t \to qh$ transitions is given by $BR(t\rightarrow qh)=\frac{\Gamma(t\rightarrow q h)}{\Gamma(t\rightarrow Total)}$). Here the decay width of the $t \to qh$ transitions is ${\Gamma}(t \to q h) = 0.78 {{\kappa}_{tqh}}^2$. In the calculations, it is assumed that the total decay width of the top quark is equal to the decay width of the $t \to Wb$ process, neglecting the light quark masses. In this analysis, the decay width of the $t \to Wb$ process, which is the primary decay channel of the top quark, is taken into account based on the results reported in the literature. Consequently, the branching ratio of the $t \to qh$ process was obtained approximately as $BR(t\rightarrow qH)=0.52{\kappa}_{tqH}^2$\cite{36,37}.
 \subsection{Cross Sections}\label{sec3}
In this presented work, the process $\mu^+\mu^- \to t\bar{q}h$ is considered for the study of BSM interactions. This process provides a rare and unique opportunity to test FCNC interactions between the Higgs boson and the top quark. 
At the beginning of the process, two oppositely charged muons ($\mu^+$ and $\mu^-$) collide and they create an intermediate state in which the Higgs boson ($h$) and the top-antiquark pair ($t\bar{q}$, $q = u, c$) are produced. In the final state, while the top quark decays via $t \to Wb$, the dominant decay mode of the SM, the W boson decays leptonically ($W \to \ell \nu$, $\ell = e, \mu$) . On the other hand, the Higgs boson decays via $H \to b\bar{b}$ to form two b-jets. Consequently, the whole process can be expressed as follows: $\mu^{+}\mu^{-} \to t\bar{q}h \to (Wb) \bar{q} (b\bar{b}) \to (\ell \nu b) \bar{q} (b\bar{b})$.

Accordingly, the Feynman diagrams, including the tree-level contributions of new physics in this process, are depicted in Figure 1. These diagrams allow more deeper understanding of the beyond-SM effects in Higgs-top quark interactions.
In this analysis, the primary background processes that have been taken into account are $tW\bar{b}$, $t\bar{t}$, $WWh$ and $WW$. These processes are similar to the signal in terms of their kinematic properties and hence, this situation complicates the decomposition process. For example, while the decay of two top quarks in the $t\bar{t}$ process produces similar particles in the final state, the $WW$ process produces a topology similar to the signal with the decays of two W bosons.
The cross section calculations for these relevant processes were carried out by simulations in the MadGraph5$\_$aMC@NLO programme \cite{31}, which is using the SM Lagrangian and within the framework of SMEFT\cite{34}. To this end, in the programme, the SM Lagrangian defined with the FeynRules package \cite {30} is extended with the Universal FeynRules Output (UFO) module\cite{33}. Thus, this configuration enables precise modelling of both signal and background phenomena.
The analysis strategy in this paper is based on suppressing backgrounds by optimising the kinematic properties of the signal. In this regard, variables particularly such as the $H \to b\bar{b}$ decay channel of the Higgs boson and the angular distances of the $b$ jets can bring out more clearly the signal/background separation. Furthermore, kinematic parameters such as the transverse momentum ($p_T$) distribution of the leptons and the invariant mass ($M_{inv}^{bb}$) ranges of the $b$ jets are expected to play a critical role in improving the sensitivity of the analysis.

In this context, the distributions of various kinematic variables for the signal and background processes ($tw\bar{b}$, $wwh$, $t\bar{t}$ and $ww$) belonging to the process $\mu^+\mu^- \to \bar{t}qh$ were analyzed for the centre of mass energy $\sqrt{s} = 3 TeV$.  All of these analyses were performed with MadAnalysis 5 and the kinematic distributions obtained ($\Delta R$, $M_{inv}^{bb}$, $M_T$, and $p_T$) are presented in Figure 2 \cite{35}. These figures show how the signal and background processes differ relative to each other in terms of different kinematic variables and provide guidance for determining appropriate cut-off points for signal/background separation.
Examining in the Figure 2, it was observed a distinctive difference in the angular distance (${\Delta R} _{b\ell}$) distributions between the $b$ jets and the positive lepton. This is due to the fact that in the signal events, one of the $b$ jets originates from the top quark decay and the other from the Higgs decay. Also, it is seen that the signal concentrates at larger values of ${\Delta R}_{b_2{\ell}}$ since the $b$ jet from the Higgs is produced angularly more distant than the lepton from the top quark. Conversely, background processes, in particular $t\bar{t}$, exhibit a more dominant behaviour in the lower $\Delta R$ ranges, since in these events the $b$ jets and leptons usually originate from the same top quark. These findings indicate that a cutoff of $\Delta R > 2.5$ can provide a significant suppression of the background while it preserves a large fraction of the signalling events. On the other hand, the invariant mass distribution of two $b$ jets ($M_{inv}^{bb}$) provides a significant advantage in distinguishing between signal and background processes. In the $t\bar{q}h$ signal, it is seen that a peak in the range of 100-130 GeV occurs due to the decay of the Higgs boson to $b\bar{b}$. This range corresponds to the characteristic decay mass of the Higgs boson and plays a critical role in the characterisation of the signal. In contrast, it is observed that there is no distinct peak in this range for the $t\bar{t}$ twb and ww background processes and the distributions are more widely spread. The findings suggest that applying a cut-off of $100 GeV <M_{inv}^{bb} < 130 GeV$ can enable the signal to be discernible. With this mind, an optimal cutoff in further analyses can be determined based on this range.
Another kinematic distribution, the transverse mass distributions ($M_T^{top}$ and $M_T^W$), provide complementary information for signal and background discrimination.  it can be said according to the $M_T^{top}$ figure, the signal is concentrated over a wide energy range and this situation reflects the kinematic properties of the top quark. On the other hand, the $M_T^W$ figure shows that the signal is more intense in lower energy ranges, which is consistent with the energy transport behaviour of the W boson. Accordingly, the relevant figures indicate that the ranges 50-100 GeV for $M_T^W$ and 100-200 GeV for $M_T^{top}$ represent the regions of highest signal concentration. This observation suggests that these ranges can be employed as cutoffs to exclude a significant portion of background processes. On the other hand, the distributions of transverse momentum ($p_T$) can provide valuable insight into the characteristics of $b$ jets and leptons. Here, in the signalling events, one of the $b$ jets, ($b_1$), comes from the top quark and the other one, ($b_2$), comes from the Higgs decay.
From the figures, it is seen that $p_T^{b_1}$ is concentrated at higher momentum values whereas $p_T^{b_2}$ is distributed to value range of the lower momentum in general. This behaviour is consistent with the fact that the $b$ jet from the Higgs decay has lower energy in general. On the other hand, for the Lepton, the signal is observed to be concentrated above 30 GeV.  In this context, in accordance with the sensitivity of the sensors in the experiments and the feasibility of the analyses, it was decided to apply cuts as $p_T^{b_1} > 30$ GeV and $p_T^\ell > 30$ GeV. Here, it can be said that the selection of these cuts is suitable for effectively suppressing low-momentum background processes while preserving the signal. Furthermore, these cuts are expected to enhance the signal/background separation under experimental conditions.

The recommended cut selection strategies to be applied in line with the findings obtained from Figure 2 are summarised in Table I for the reader to easily see the cuts. As it is seen in Table I, the limits of $p_T^\ell > 10$ GeV and $p_T^j > 20$ GeV for leptons and jets, respectively, were set as the minimum cuts (‘basic cut’) to form the basis for the analyses. Then, basic kinematic selections like $|\eta^{b_1,b_2,\ell,j}| < 2.5$ for $E_T^{\text{missing}} > 30$ GeV, $p_T^{b_1,b_2,\ell,j} > 30$ GeV and $\eta$ variables were gradually added (Cut1).  Next, cuts such as $\Delta R_{{b_2}\ell} > 2. 5$ (Cut2), $50 \text{ GeV} < M_T^W < 100 \text{ GeV}$ (Cut3), $100 \text{ GeV} < M_T^{\text{top}} < 200 \text{ GeV}$ (Cut4) and finally $100 \text{ GeV} < M_{inv}^{bb} < 130 \text{ GeV}$ (Cut5) were applied. This cuts order sequence was designed to ensure signal conservation and efficiently background suppression. In a summary, these analyses suggest that the signal can be more effectively isolated when multivariate cut selection strategies are used. Therefore, combining these kinematic variables to distinguish the signal from the background is a strategy to increase the sensitivity of the analysis. Moreover, the gradual implementation of the cuts is another critical strategy to increase the analysis sensitivity and optimise the isolation of the signal.
The findings obtained so far have allowed us to understand the behaviours of different kinematic properties of the signal and background processes and have, hence, enable to identify effective cut strategies. At this point, in order to quantitatively evaluate the effect of the chosen cut  strategies on the success of the analyses, the cross sections of the signal and SM background processes  using the determined cuts for the $\mu^+\mu^- \to \bar{t}qh$ process were calculated at two different centre-of-mass energies ($\sqrt{s} = 1 {TeV}$ and $\sqrt{s} = 3 {TeV}$) . Table II and Table III list the cross sections and signal-to-background ratios (S/B) obtained at these two energy values, respectively. Here, these tables were presented to evaluate the effect of the gradual application of the determined cuts on the isolation of the signal.
Table II represents analyses a $\sqrt{s} = 1{TeV}$ energy level . At the basic cut stage, the signal cross section is $2.91 \times 10^{-2} {pb}$, while the total background is $1.587 \times 10^{-1} {pb}$. The B/W ratio at this stage is 0.17. After the application of Cut1, it is seen that the cross-sections of both signal and background decreased, however the $S/B$ ratio increased slightly ($0.18$). When further cuts (Cut2 and Cut3) were applied, the cross-section of the background processes was further suppressed, but the S/B ratio decreased significantly at Cut3 ($0.10$). The most remarkable results were obtained at Cut4 and Cut5. After these cuts, $t\bar{t}$ and other background processes were effectively eliminated and it was resulted with a $S/B$ ratio of $49.1$ and $2.9\times 10^3$, respectively. As a consequence, these finding indicates that the $\Delta R$, $M_T$, and $M_{inv}^{bb}$ cuts play a critical role in background suppression.

Analyses on $\sqrt{s} = 3 TeV$ energy level is listed Table II. At this energy level, the signal cross section at the basic cut stage is $3.72 \times 10^{-3}{pb}$, while the total background is calculated to be $8.246 \times 10^{-3}{pb}$. This situation compared with $\sqrt{s} = 1 {TeV}$ shows that the signal cross section starts at a lower value, but the S/B ratio is much higher ($0.41$). When the cuts are applied gradually, especially from Cut2 onwards, it is observed that the $t\bar{t}$ background process is completely suppressed. Also, it is clearly seen in the table that in Cut4 and Cut5, other background processes were also significantly reduced and eventually Cut5 resulted in an impressive B/W ratio of $7.3 \times 10^3$. These results reveal that $\sqrt{s} = 3 TeV$ energy level provides a clear advantage in terms of signal isolation and suppression of background processes.
Evaluating the effect of cuts in terms of different background processes, it is reavel that some cuts are particularly effective in suppressing certain backgrounds. To illustrate, the $\Delta R > 2.5$ cut applied in Cut2 is particularly effective in suppressing $t\bar{t}$ processes since in these processes, $b$ jets and leptons usually originate from the same top quark and are concentrated at lower $\Delta R$ values. Similarly, it is observed that the cut $50 \text{ GeV} < M_T^W < 100 \text{ GeV}$ applied in Cut3 targets the transverse mass distribution of processes like $wwh$ and $ww$ and thereby suppressing these backgrounds to a  significant large extent. Additionally, it is observed that the $M_T^{\text{top}}$ and $M_{inv}^{bb}$ cuts applied in Cut4 and Cut5 effectively suppress both the $t\bar{t}$ process and the $tw\bar{b}$ background. In particular, the $M_{inv}^{bb}$ cut targets the characteristic $b\bar{b}$ decay mass range of the Higgs boson and it largely excludes background processes distributed outside this region.
When we evaluate both tables together, it is evident that $\sqrt{s} = 3 {TeV}$ energy level provides a distinct advantage in terms of signal/background separation and analysis sensitivity. In particular, the $S/B$ ratio achieved at the Cut5 stage indicates that the signal can be almost completely isolated at this energy level.  At $\sqrt{s} = 3 TeV$, the $t\bar{t}$ process is completely suppressed from Cut2 onwards and the cross sections of the other background processes decrease to very low values as a result of the application of the Cut5 cut. This situation emphasises the analytical power of this energy level.
As a result, the analyses at $\sqrt{s} = 3 TeV$, reveal that the determined cuts provide a more effective distinguishing of the signal and a more successful suppression of the background. It can thus be stated that this energy level presents a substantial opportunity for the isolation of the signal and the enhancement of the sensitivity of the analysis. These findings indicate that the preference for the energy level of $\sqrt{s} = 3 TeV$ energy level in the muon collider will optimise the performance of the analysis. 
In the analyses conducted thus far, the use the kinematic variables for the process $\mu^+\mu^- \to t\bar{q}h$ has enabled the isolation of the signal and the suppression of the background processes.

\section{Sensitivity Analysis}

Following the above systematic cut selection strategies, in order to more deeper assess the detectability of the process, the relationship between branching ratios ($\text{BR}(t \to qH)$) and limunosity ($\mathcal{L}$) was investigated via the statistical significance test of $Z_{excl}$. Here, $Z_{excl}$ is a statistical metric that evaluates the distunguishibility of the signal over the background and is also critical for determining the experimental exclusion limits in the analyses.  $Z_{excl}$ is defined as follows \cite{38}: 

\begin{eqnarray}
Z_{excl}=\sqrt{2[s-b\ln(\frac{s+b+x}{2b})-\frac{1}{{\delta}^2}\ln(\frac{b-s+x}{2b})]-(b+s-x)(1+\frac{1}{{\delta}^{2}b})}
\end{eqnarray}

In this equation, s represents the expected number of signal events, b denotes the expected number of background events and $ \delta$ is the systematic error parameter. Here, $ x=\sqrt{(s+b)^{2}-(\frac{4{\delta}^{2}sb^{2}}{1+{\delta}^{2}b}})$ is an optimised correction term of the systematic uncertainty.
When the effect of the systematic error parameter ($\delta$) is negligible ($\delta \to 0$), $Z_{excl}$ can be expressed in a simplified form:

 \begin{equation} 
Z_{excl} = \sqrt{2 \left[ s - b \ln\left(1 + \frac{s}{b}\right) \right]}.
 \end{equation}
 
$Z_{excl}$ is used to calculate how far the signal can be distinguish from background processes in the statistical significance assessment. Additionally, the $Z_{excl}$ value enables the assessment of the sensitivity of the analyses by combining the detectability of the signal and the impact of systematic error. In this regard, for the limit of detectability at the $95\%$ confidence level, $Z_{excl} \ge 1.65$ is employed. This is a key role to ascertain that the analyses present meaningful results and not merely coincidental.
To this end, the calculations are performed based on the light of this information, both $\sqrt{s} = 1 {TeV}$ and $\sqrt{s} = 3 TeV$ energy levels considering two different systematic errors ($\delta = 0\%$ and $\delta = 10\%$) and the obtained results are presented in Figure 3. In the figure, top panel is for the $\sqrt{s} = 1TeV$ energy level analyses while the bottom panel is for the$\sqrt{s} = 3 TeV$ ones.

An examination of Figure 3 reveals that at the energy level of $\sqrt{s} = 1 {TeV}$, the branching ratio decreases rapidly with increasing luminosity. In this case, it can be said that when the systematic error is $\delta = 0\%$, in order to achieve the ATLAS experimental limit $\mathcal{L} \approx 8 {ab}^{-1}$ integrated luminosity is required. In the event of a systematic error $\delta = 10\%$, however, the required luminosity value for the limit obtained in our calculations to remain below the ATLAS experimental limit is $\mathcal{L} \ge 20 {ab}^{-1}$. For example, at the highest calculated luminosity value $\mathcal{L} = 20 {ab}^{-1}$, we observe that the branching ratio decreases to $3.4 \times 10^{-4}$ for $\delta = 10\%$. Accordingly, this result is insufficient to improve the current experimental limits. This situation furthermore corresponds that the effect of systematic errors is more pronounced, especially at low radii, limiting the precision of the analysis. On the other hand, at the relevant centre-of-mass energy and luminosity range, the lowest limit obtained for $\text{BR}(t \to qH)$ is about $1.96 \times 10^{-4}$ when $\delta = 0\%$.  This value is about 1.5 times more restrictive than the latest experimental limits obtained in ATLAS.
According to the analyses conducted at $\sqrt{s} = 3 {TeV}$ energy level, it is observed that the systematic errors do not cause fluctuations and therefore the results of the $\delta = 0\%$ and $\delta = 10\%$ are very close to each other.  As is evident from the figure, the branching ratios in both $\delta = 0\%$ and $\delta = 10\%$ remain below the ATLAS experimental limit and these ratios exhibit a rapid decline with the increase in luminosity. Here, as an striking finding, it is calculated that the required luminosity to achieve the ATLAS limit, independent of the systematic error ($\delta$) level, is approximately $\mathcal{L} \approx 4 {ab}^{-1}$. This situation indicates that the effect of systematic error does not make a significant difference at high energy levels and the sensitivity of the analysis is remarkably consistent. Furthermore, $\sqrt{s} = 3 {TeV}$ energy level, the tightest limit obtained for $\text {BR}(t \to qH)$ is $1.1 \times 10^{-4}$. This value represents about a 3-times improvement compared to the most recent experimental limits obtained in ATLAS. This finding clearly demonstrates that this energy level provides a highly sensitive analysis environment for signal decomposition as independent of the systematic error level.
When the branching ratio figures obtained for two different centre of mass energies are scrutinized together, the analytical superiority at $\sqrt{s} = 3 TeV$ is clearly obvious.  For example, The branching ratios at $\mathcal{L} = 5 {ab}^{-1}$  level was calculated as  $\text{BR}(t \to qH) \approx 4.1 \times 10^{-4}$  for $\sqrt{s} = 1 TeV$ and $\text{BR}(t \to qH) \approx 2.8 \times 10^{-4}$  for $\sqrt{s} = 3 TeV$. This analysis reveals that the $\sqrt{s} = 3 TeV$ energy level significantly reduces the effect of systematic error by enhancing the detection sensitivity.
Consequently, the analyses presented in Figure 3 suggest that $\sqrt{s} = 3 TeV$ energy level exhibits a structure more sensitive to systematic errors and optimises the detectability of the $\mu^+\mu^- \to t\bar{q}h$ process. Moreover, the $\sqrt{s} = 3TeV$ energy level allows for precise analysis even at lower luminosities and enables the lowest limits in terms of branching ratio to be achieved at lower levels. Therefore, it is a highly powerful option for use in a muon collider.

\section{Conclusion}
This study investigates the FCNC interactions between the Higgs boson and the top quark by examining the process $\mu^+\mu^- \to t\bar{q}h$ in the SMEFT framework. At this juncture, by exploiting the advantages of muon colliders such as high energy efficiency and clean collision environment, the potential of this rare process in the search for BSM physics is evaluated.
The analyses revealed that kinematic variables such as $\Delta R$, $M_{inv}^{bb}$, $M_T$ and $p_T$ play a critical role in distinguishing the signal from background processes such as $t\bar{t}$, $tW\bar{b}$, $WWWh$ and $WW$. Accordingly, the application systematically of optimised cuts leads to a significant improvement in the signal-to-background (S/B) ratios, particularly at $\sqrt{s} = 3$ TeV energy level. In this context, the $\Delta R$ and $M_{inv}^{bb}$ cuts are found to preserve a large fraction of the signal events while they effectively suppress background processes.
Furthermore, the evaluation of the relationship between the branching ratio ($\text{Br}(t \to qH)$) and the luminosity ($\mathcal{L}$) reveal that analyses at $\sqrt{s} = 3$ TeV energy level provide superior sensitivity. For example, the branching ratio limit for $\mathcal{L} = 5 ab^{-1}$ is calculated to be $2.9 \times 10^{-4}$ , which goes slightly beyond the current limits of the ATLAS experiment. At higher luminosity and neglecting systematic errors, $\sqrt{s} = 3$ TeV, the stringent limit obtained for $\text{Br}(t \to qH)$ is $1.1 \times 10^{-4}$. This value represents an improvement of approximately 3 times compared to the $3.0 \times 10^{-4}$ limit provided by the ATLAS experiment. These results demonstrate the potential of rare phenomena beyond the SM to extend the limits of observability at muon colliders. Moreover, the effect of systematic uncertainties is observed to diminish at higher centre-of-mass energies, and hence, it is concluded that the $\sqrt{s} = 3$ TeV energy level in particular can provide an ideal platform for precise analyses and for exploring previously inaccessible frontiers of FCNC processes.
In conclusion, the findings of the study have revealed the potential of muon colliders in the sensitive search for rare processes such as $\mu^+\mu^- \to t\bar{q}h$ and will be an important step towards understanding FCNC interactions. Therefore, these results are expected to shed light on future theoretical and experimental studies in a more precise search for traces of new physics.

\begin{figure}
\includegraphics[width=18.cm]{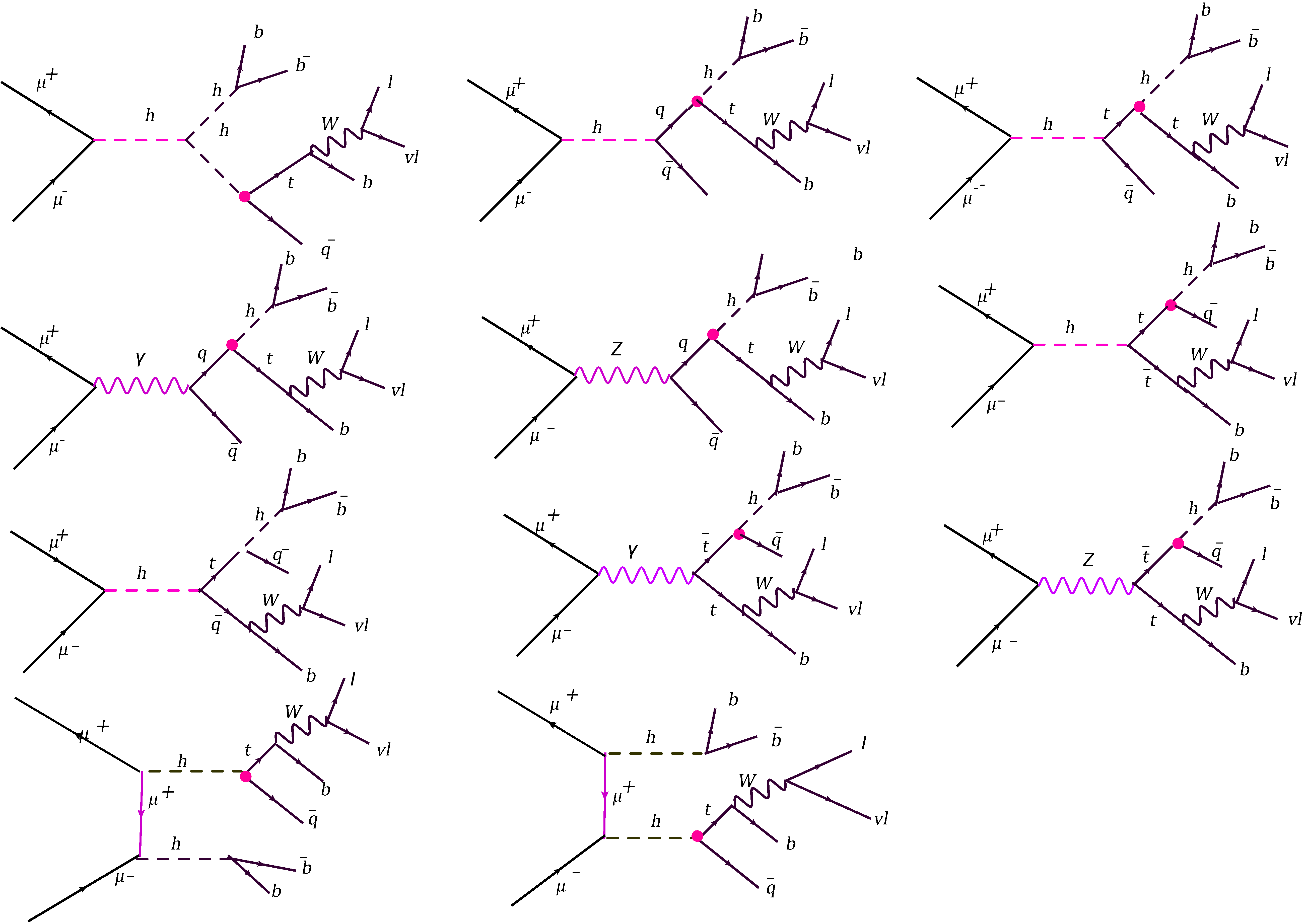}
\caption{Tree level Feynman diagrams for the  $\mu^{+}\mu^{-} \to t\bar{q}h \to (Wb) \bar{q} (b\bar{b}) \to (\ell \nu b) \bar{q} (b\bar{b})$ process that includes new physics contributions.}
\end{figure}

\begin{figure}[htp]
    \centering
    \begin{subfigure}[b]{0.45\textwidth}
        \centering
        \includegraphics[width=\textwidth]{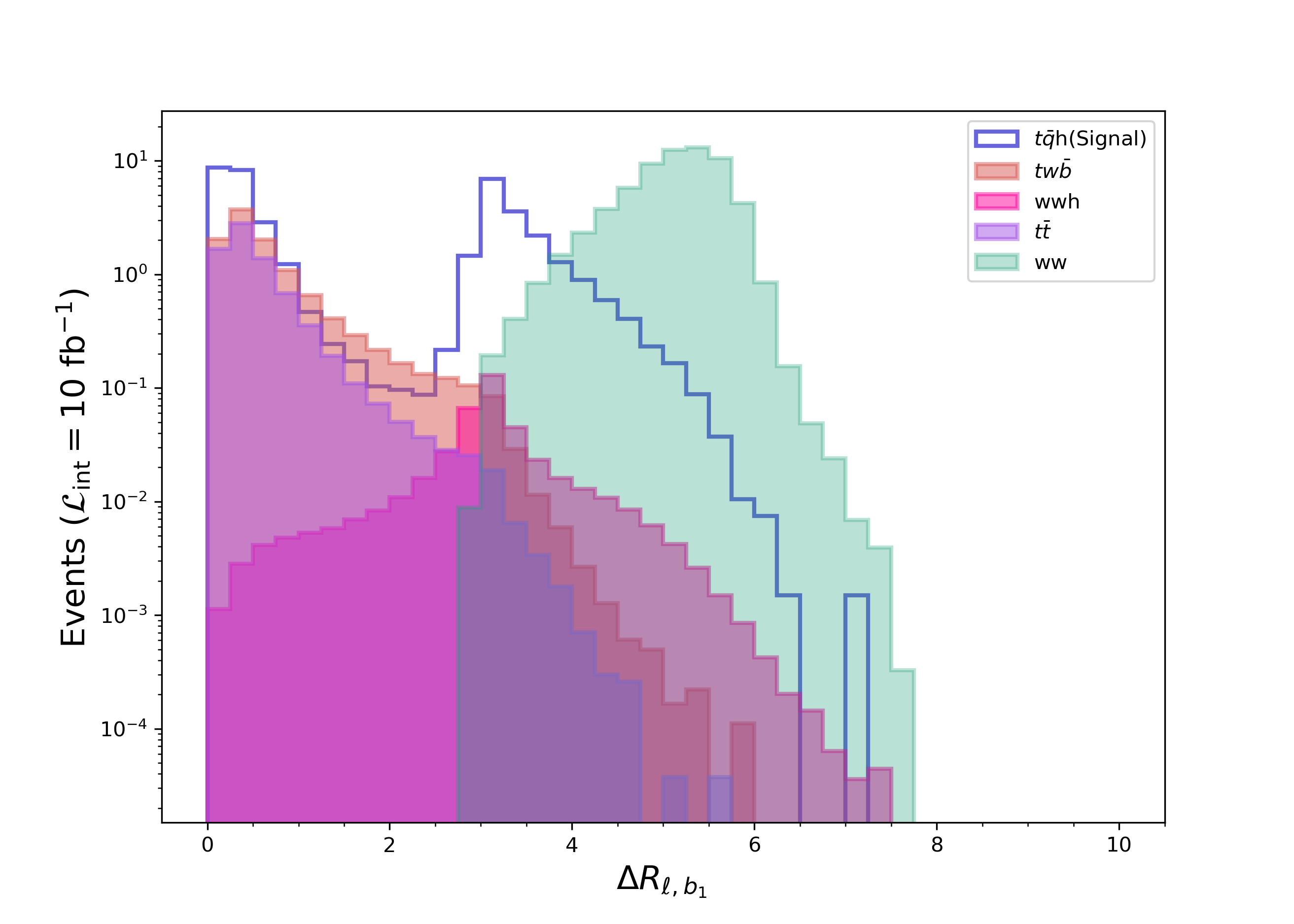} 
    \end{subfigure}
    \hspace{0.5cm}
    \begin{subfigure}[b]{0.45\textwidth}
        \centering
        \includegraphics[width=\textwidth]{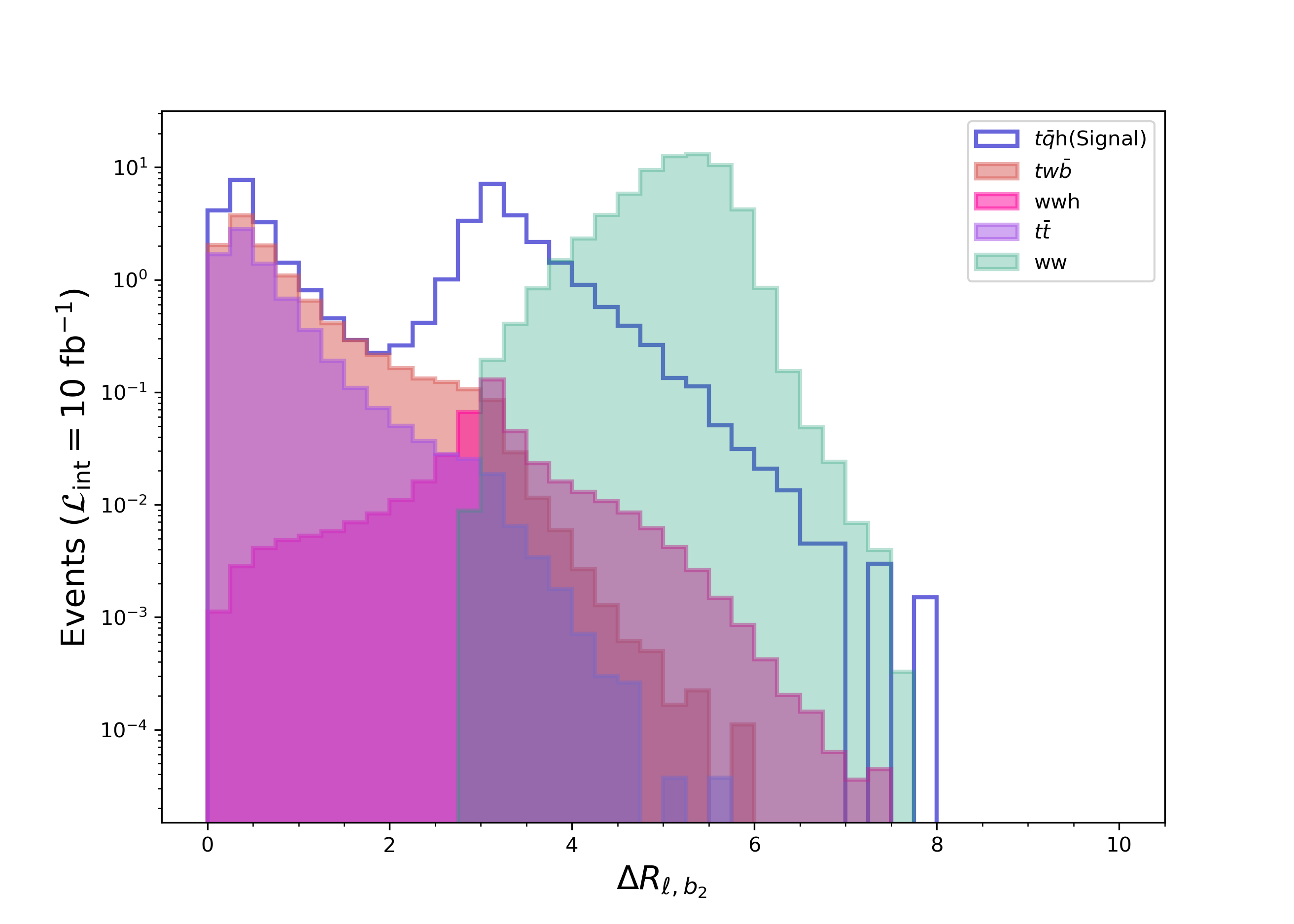} 
    \end{subfigure}
    \vspace{0.5cm}
    \begin{subfigure}[b]{0.45\textwidth}
        \centering
        \includegraphics[width=\textwidth]{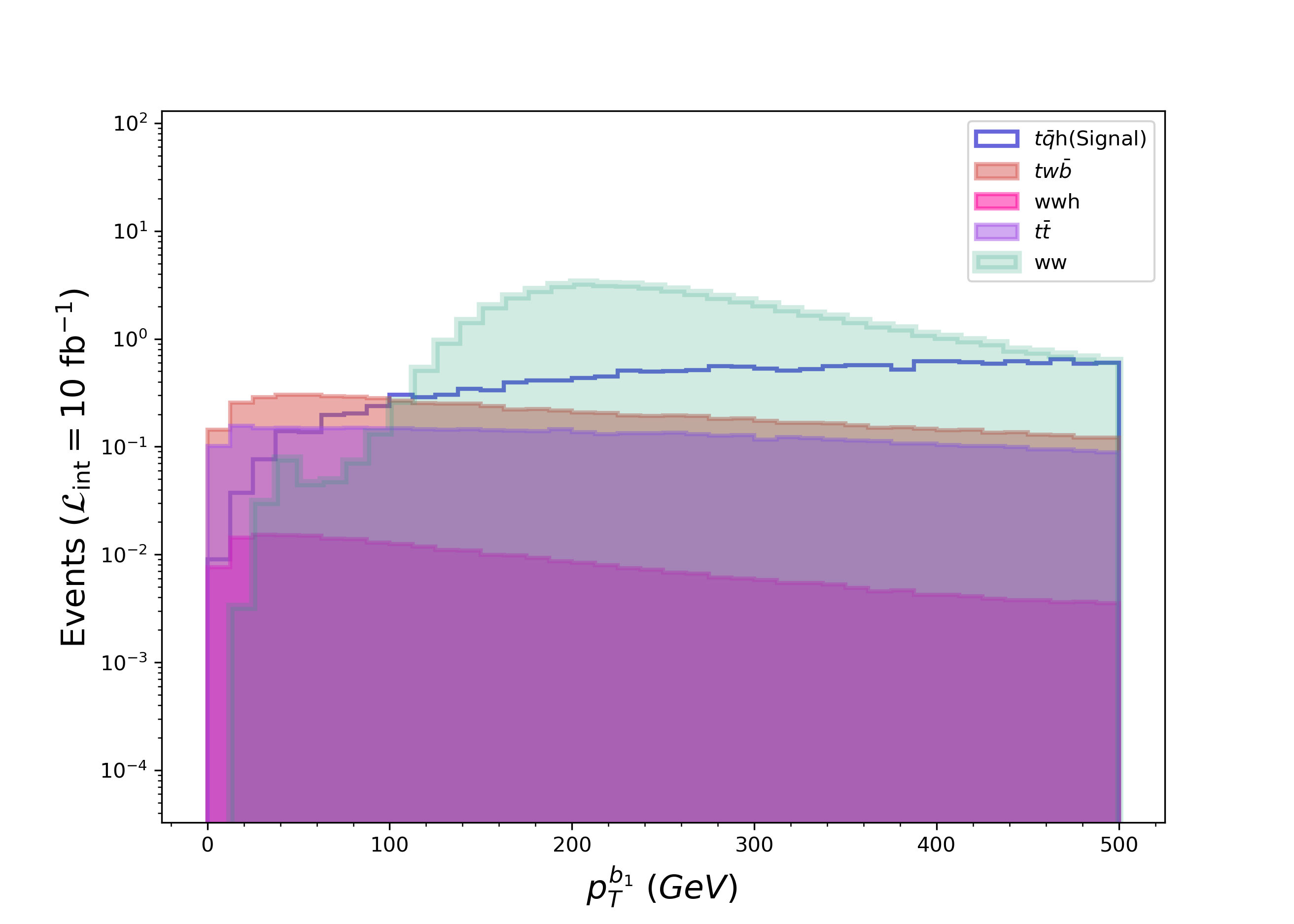} 
    \end{subfigure}
    \hspace{0.5cm}
    \begin{subfigure}[b]{0.45\textwidth}
        \centering
        \includegraphics[width=\textwidth]{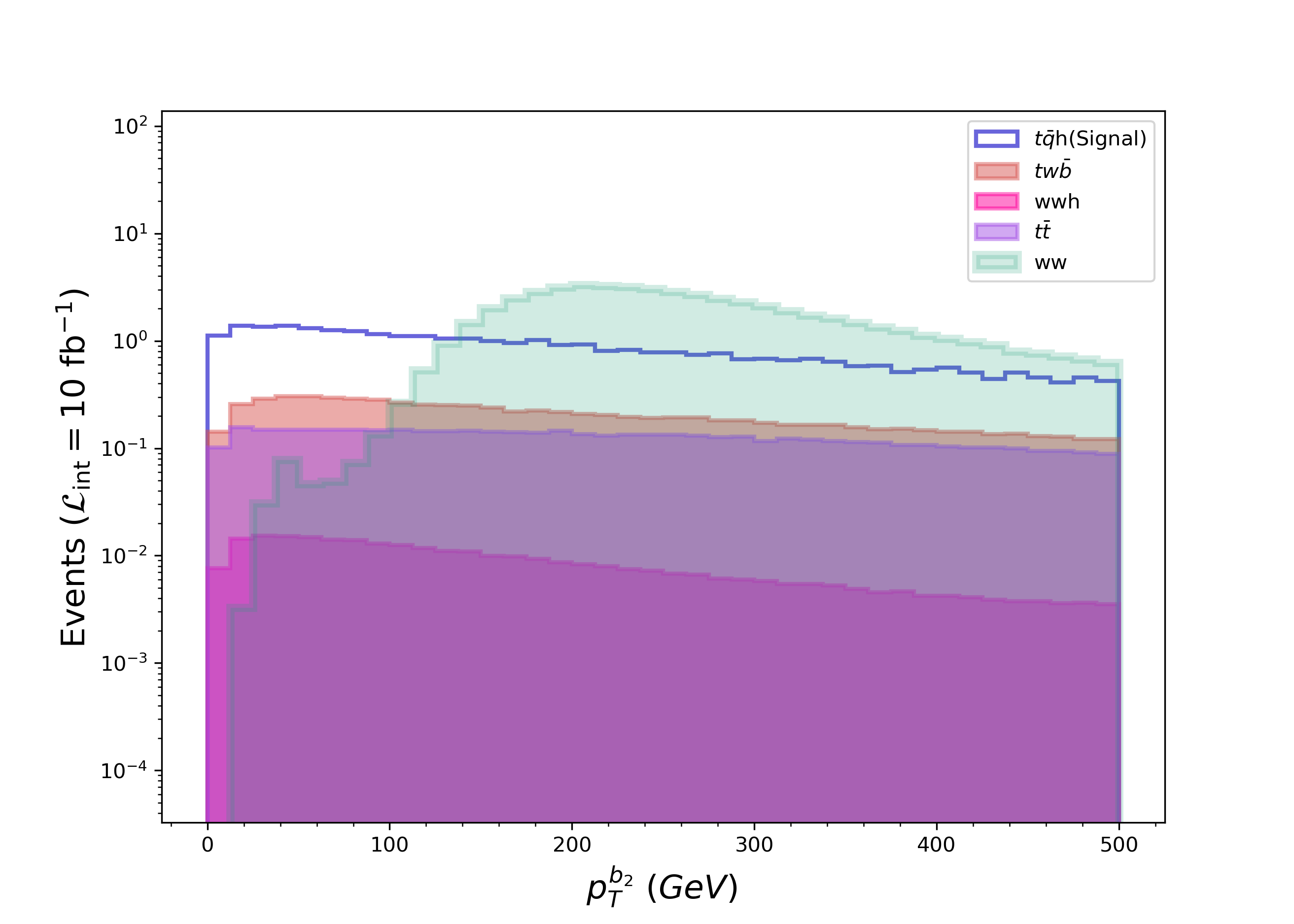} 
    \end{subfigure}
            \vspace{0.5cm}
    \begin{subfigure}[b]{0.45\textwidth}
        \centering
        \includegraphics[width=\textwidth]{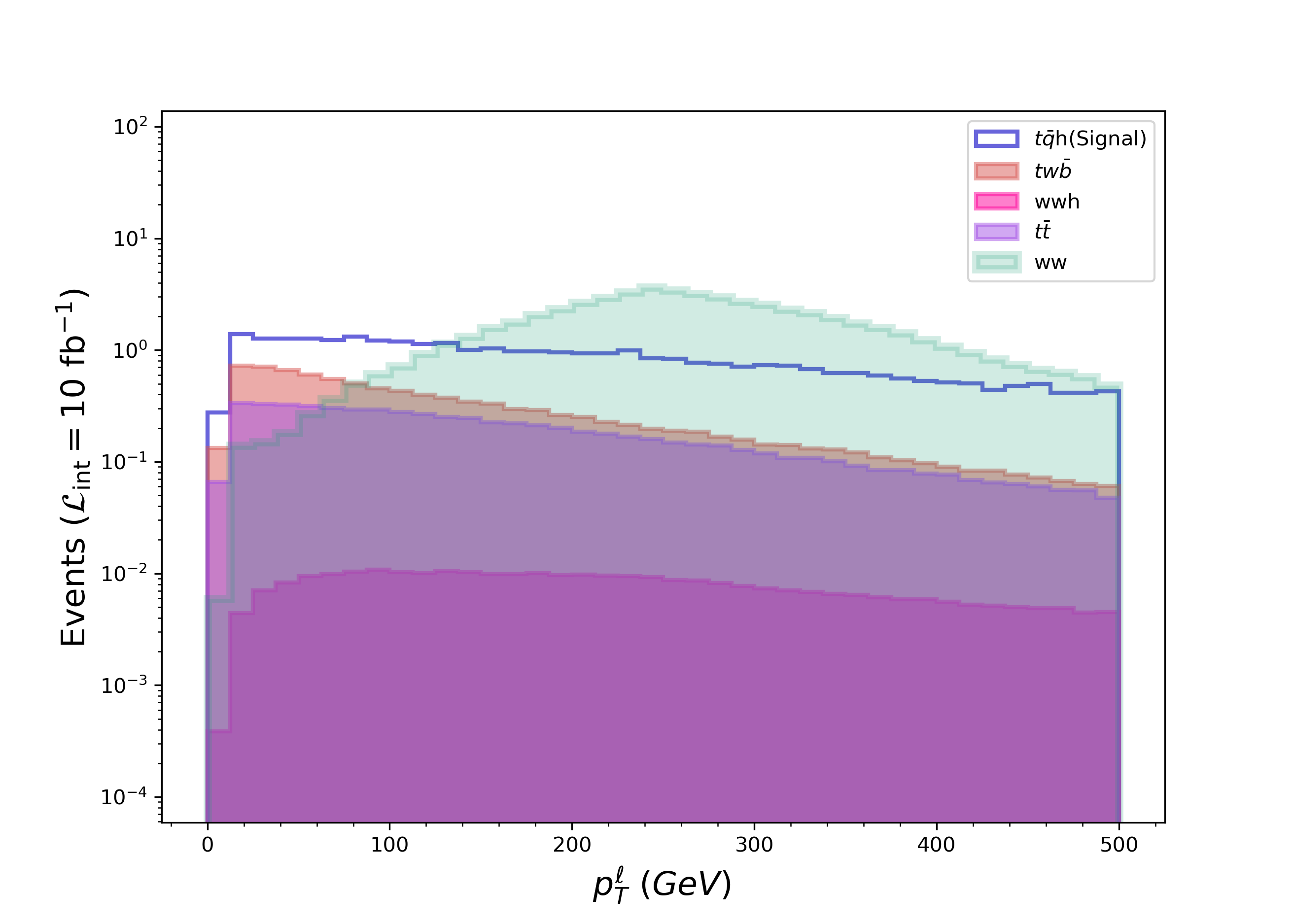} 
    \end{subfigure}
     \hspace{0.5cm}
    \begin{subfigure}[b]{0.45\textwidth}
        \centering
        \includegraphics[width=\textwidth]{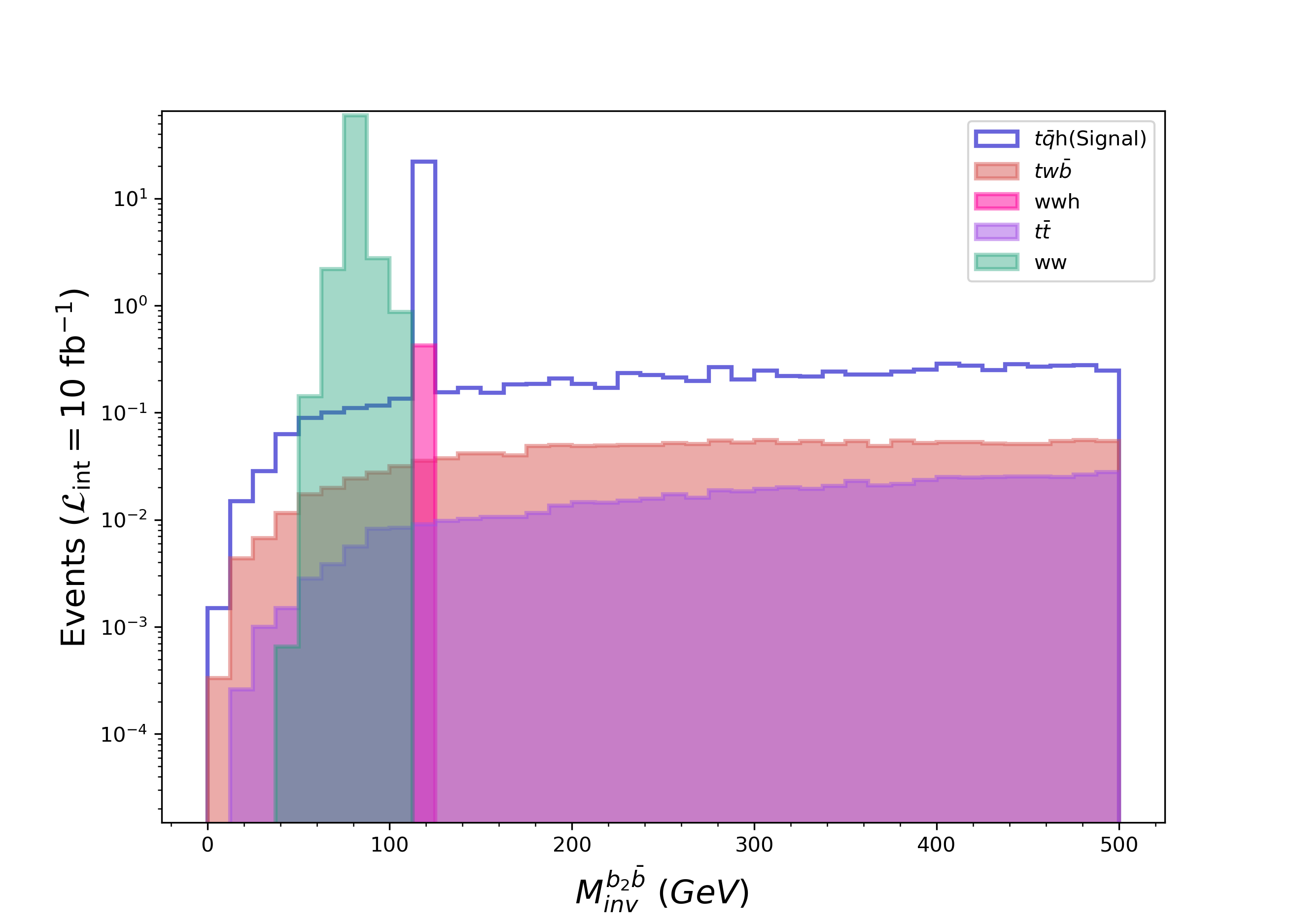} 
    \end{subfigure}
    \vspace{0.5cm}
    \begin{subfigure}[b]{0.45\textwidth}
        \centering
        \includegraphics[width=\textwidth]{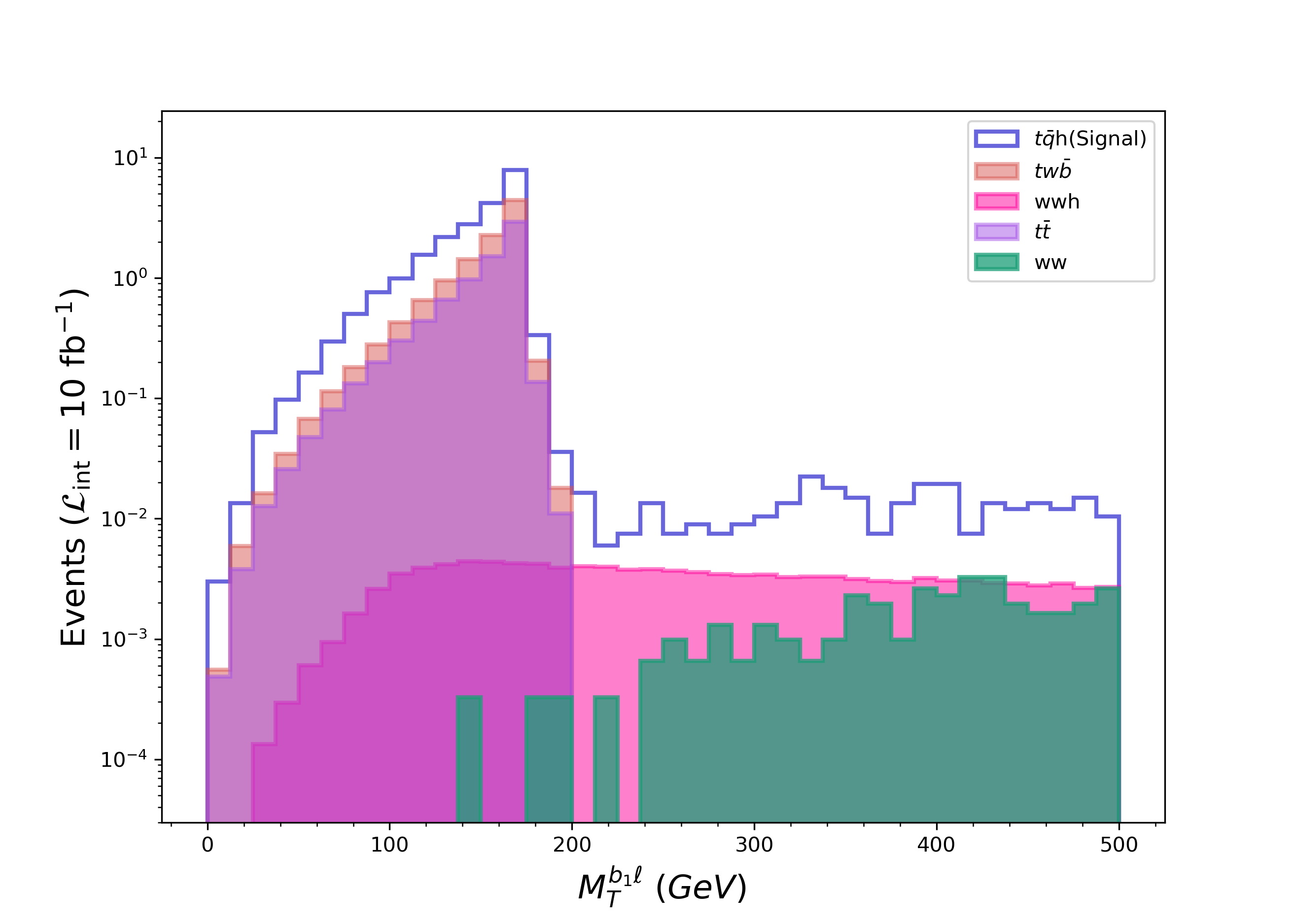} 
    \end{subfigure}
     \hspace{0.5cm}
    \begin{subfigure}[b]{0.45\textwidth}
        \centering
        \includegraphics[width=\textwidth]{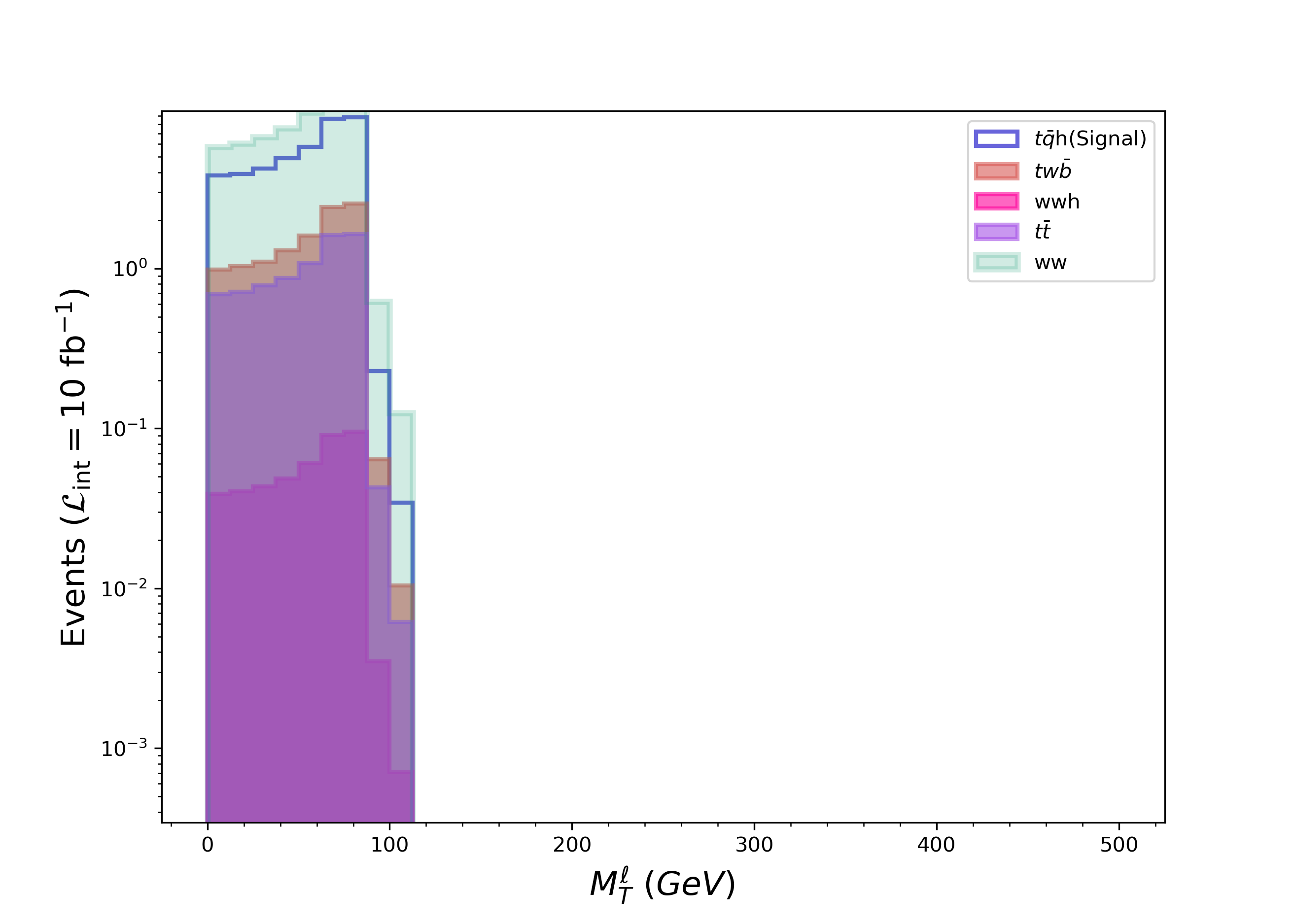} 
    \end{subfigure}
    
    \caption{Some kinematical distributions for the signals($t \bar{q} h$) and SM backgrounds ($W b \gamma$ , $W j \gamma$ ,$ Z j \gamma$)  for center-of mass energy of $\sqrt{s}=3TeV$ at the MuC.}
\end{figure}

\begin{table}[h]
\begin{center}
\caption{List of applied kinematic cuts for the analysis.}
\begin{tabular}{cccc|cccc}
\hline\hline
Basic cut& $ p_{T}^{\ell}> 10 GeV$, $ p_{t}^{j}> 20 GeV$ &\\
Cut1&  $E_{T}^{missing}> 30$ GeV, $p_{T}^{b_{1},b_{2},\ell}> 30$ GeV,$\left|{\eta}^{b_{1},b_{2},\ell,j}\right|<2.5 $,  & \\
Cut2& Same as cut 1 +${\Delta R}_{{b_2}\ell}>2.5$   & \\
Cut3&  cut2 + $50 GeV< M_{T}^W<100 GeV$ &  \\
Cut4&  cut3 + $100 GeV< M_{T}^{top}<200 GeV$ & \\
Cut5&  cut4 + $100 GeV< M_{inv}^{bb}<130 GeV$ & \\
\hline\hline
\end{tabular}
\end{center}
\end{table}

\begin{table}
\caption{The cut-based analysis of the cross sections (in pb) for signal and SM background events  at the muon collider with $\sqrt{s}=1 TeV$  assuming  $\kappa_{q}=1$.}
\begin{center}
\begin{tabular}{c|c|c|c|c|c|c}
\hline\hline
\multicolumn{6}{c}{$\sqrt{s}=1 TeV$} \\
\hline\hline
\cline{1-5}
\multicolumn{1}{c|}{}&\multicolumn{1}{c|}{ signal} &\multicolumn{3}{l}{Background}\\
\hline
cuts&$\mu^{+}\mu^{-}\rightarrow t\bar{q}h$&$t\bar{t}$ &$ twb$ &$ wwh $  &$ ww $&S/$B_{tot}$ \\
\hline
Basic cut  &$2.91\times10^{-2}$   &$1.557\times10^{-2}$ &$1.647\times10^{-2}$&$3.409\times10^{-4}$&$1.431\times10^{-1}$&0.17   \\
\hline
cut1  &$1.346\times10^{-2}$  &$7.272\times10^{-3}$&$8.26\times10^{-3}$&$1.37\times10^{-4}$&$5.99\times10^{-2}$& 0.18  \\
cut2 &$5.787\times10^{-3}$   & $7.32\times10^{-5}$  &$1.10\times10^{-4}$&$8.14\times10^{-5}$&$5.98\times10^{-2}$& 0.10  \\
cut3  &$3.604\times10^{-3}$   & $5.72\times10^{-5}$          &$8.44\times10^{-6}$&$5.34\times10^{-5}$&$3.70\times10^{-2}$&0.10   \\
cut4  &$3.504\times10^{-3}$   &$5.72\times10^{-5}$         &$8.44\times10^{-6}$&$4.87\times10^{-6}$&$7.2\times10^{-7}$&49.1   \\
cut5  &$3.504\times10^{-3}$   &$2.31\times10^{-6}$         &$4.67\times10^{-6}$&$4.87\times10^{-6}$&$2.4\times10^{-7}$&$2.9\times10^3$   \\
\hline
\end{tabular}
\end{center}
\end{table}

\begin{table}
\caption{The cut-based analysis of the cross sections (in pb) for signal and SM background events  at the muon collider with $\sqrt{s}=3 TeV$  assuming  $\kappa_{q}=1$.}
\begin{center}
\begin{tabular}{c|c|c|c|c|c|c}
\hline\hline
\multicolumn{6}{c}{$\sqrt{s}=3 TeV$} \\
\hline\hline
\cline{1-5}
\multicolumn{1}{c|}{}&\multicolumn{1}{c|}{ signal} &\multicolumn{3}{l}{Background}\\
\hline
cuts&$\mu^{+}\mu^{-}\rightarrow t\bar{q}h$&$t\bar{t}$ &$ twb$ &$ wwh $  &$ ww $ &S/$B_{tot}$\\
\hline
Basic cut  &$3.72\times10^{-3}$   &$7.984\times10^{-4}$ &$1.11\times10^{-3}$&$4.695\times10^{-5}$&$7.09\times10^{-3}$&0.41  \\
\hline
cut1  &$2.84\times10^{-3}$   &$5.47\times10^{-4}$&$6.92\times10^{-4}$&$2.02\times10^{-5}$&$3.71\times10^{-3}$&0.57   \\
cut2 &$1.51\times10^{-3}$   & 0 &$1.27\times10^{-6}$&$1.84\times10^{-5}$&$3.71\times10^{-3}$&0.40  \\
cut3  &$8.79\times10^{-4}$   & 0 &$9.98\times10^{-7}$&$1.07\times10^{-6}$&$2.21\times10^{-3}$&0.40   \\
cut4  &$8.67\times10^{-4}$   & 0 &$9.98\times10^{-7}$&$9.66\times10^{-8}$&0&$792$  \\
cut5  &$8.67\times10^{-4}$   & 0 &$2.08\times10^{-8}$&$9.66\times10^{-8}$&0&$7.3\times10^3$  \\
\hline
\end{tabular}
\end{center}
\end{table}

\begin{figure}[htp]
    \centering
    \begin{subfigure}[b]{0.75\textwidth}
        \centering
        \includegraphics[width=\textwidth]{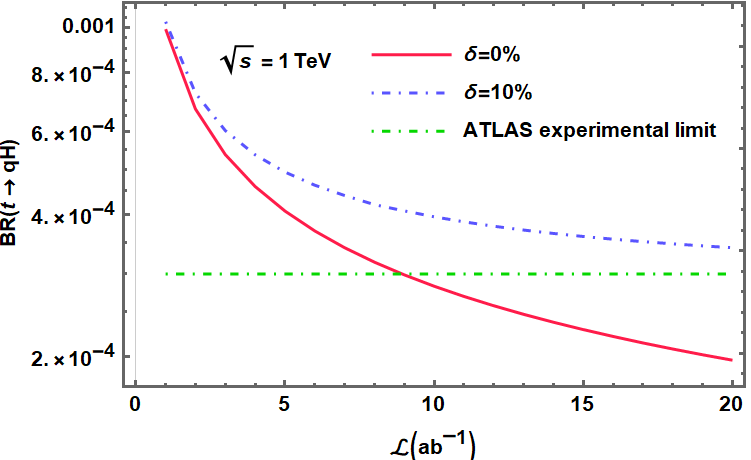} 
    \end{subfigure}
    \hspace{0.5cm}
    \begin{subfigure}[b]{0.75\textwidth}
        \centering
        \includegraphics[width=\textwidth]{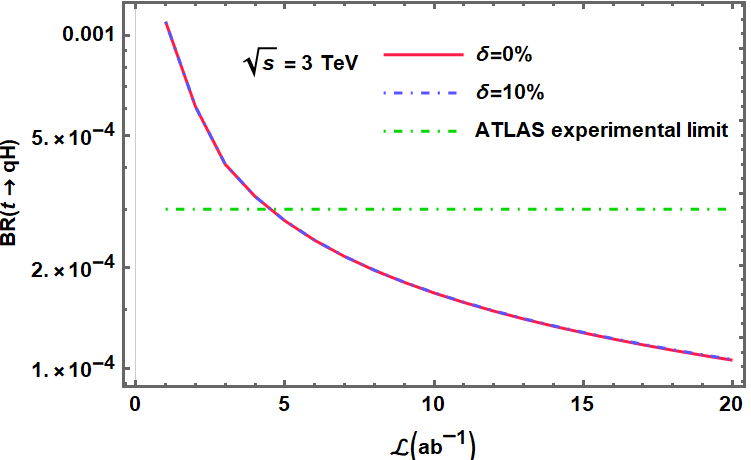} 
    \end{subfigure}
        \caption{95\% C.L. bounds on the $\text{Br}(t \to qh)$ for center of mass energy $\sqrt{s}=1.0$ TeV and $\sqrt{s}=3.0$ TeV through the processes $\mu^{+}\mu^{-} \to t\bar{q}h \to (Wb) \bar{q} (b\bar{b}) \to (\ell \nu b) \bar{q} (b\bar{b})$.}
\end{figure}

\end{document}